\documentclass[sigconf]{acmart}

\usepackage{microtype}
\usepackage{xcolor}
\usepackage{xspace}

\newcommand{\eg}{e.g.,\xspace}
\newcommand{\eat}[1]{}
\newcommand{\paratitle}[1]{\noindent\textbf{#1}\ \ }

\AtBeginDocument{%
  \providecommand\BibTeX{{%
    \normalfont B\kern-0.5em{\scshape i\kern-0.25em b}\kern-0.8em\TeX}}}

\copyrightyear{2023} 
\acmYear{2023} 
\setcopyright{rightsretained} 
\acmConference[WWW '23 Companion]{Companion Proceedings of the ACM Web Conference 2023}{April 30-May 4, 2023}{Austin, TX, USA}
\acmBooktitle{Companion Proceedings of the ACM Web Conference 2023 (WWW '23 Companion), April 30-May 4, 2023, Austin, TX, USA}
\acmDOI{10.1145/3543873.3587309}
\acmISBN{978-1-4503-9419-2/23/04}

\begin{document}

\title{\includegraphics[height=1em]{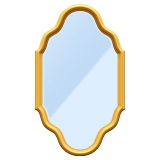}Mirror: A Natural Language Interface for Data Querying, Summarization, and Visualization} %Plug-and-Play Data Query, Summarization and Visualization with a Natural Language Interface}

\author{Canwen Xu}
\orcid{0000-0002-1552-999X}
\affiliation{%
  \institution{University of California, San Diego}
  \state{California}
  \country{USA}
}
\email{cxu@ucsd.edu}

\author{Julian McAuley}
\affiliation{%
  \institution{University of California, San Diego}
  \state{California}
  \country{USA}
}
\email{jmcauley@eng.ucsd.edu}

\author{Penghan Wang}
\authornote{To whom correspondence should be addressed.}
\affiliation{%
  \institution{Cisco}
  \state{California}
  \country{USA}
}
\email{penghawa@cisco.com}

% \renewcommand{\shortauthors}{Xu, McAuley and Wang}

\begin{abstract}
% Original abs

% We propose Mirror, a language model-powered open-source platform for data exploration and analysis. We provide an easy-to-use natural language interface to automatically query the database. By generating executable SQL command, Mirror can retrieve relevant data and automatically summarize it in natural language. Mirror also allows users to preview the generated SQL commands and manually edit them if necessary. Mirror is also capable of automatically generating visualization to better illustrate the results. With its flexible and human-in-the-loop design, Mirror can help people better understand their data, no matter they are experienced data analyst or marketing professionals without any programming knowledge.

% Revised by ChatGPT
We present Mirror, an open-source platform for data exploration and analysis powered by large language models. Mirror offers an intuitive natural language interface for querying databases, and automatically generates executable SQL commands to retrieve relevant data and summarize it in natural language. In addition, users can preview and manually edit the generated SQL commands to ensure the accuracy of their queries. Mirror also generates visualizations to facilitate understanding of the data. Designed with flexibility and human input in mind, Mirror is suitable for both experienced data analysts and non-technical professionals looking to gain insights from their data.\footnote{The code is available at \url{https://github.com/mirror-data/mirror}.}
\end{abstract}



%%
%% . The author(s) should pick words that accurately describe



\maketitle

\section{Introduction}

% \begin{figure*}[t]
%   \centering
%   \includegraphics[width=0.85\linewidth]{annotated_screenshot_mirror.jpg}
%   \caption{\label{fig:mirror}The annotated user interface of Mirror.}
% \end{figure*}

\begin{figure*}[t]
  \centering
  \includegraphics[width=0.76\linewidth]{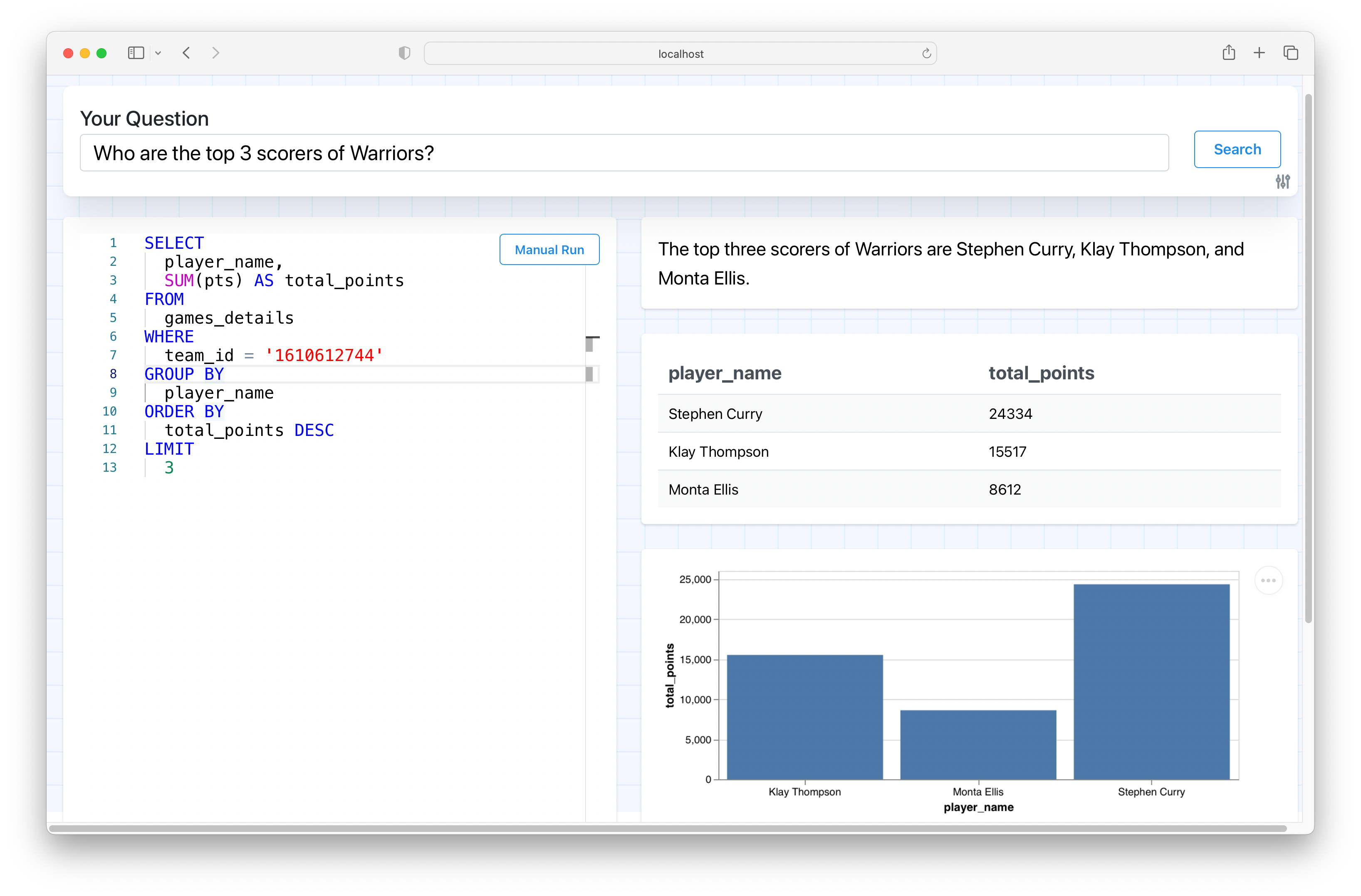}
  \vskip -1.5em
  \caption{\label{fig:mirrorui}The user interface of Mirror. \textit{(Top)} The input field allows the user to type in any query in natural language. It can suggest auto-completion to user input based on metadata. \textit{(Bottom left)} The code editor displays SQL command generated by the pretrained code model. It allows users to edit and re-run the SQL command. \textit{(Bottom right)} The automatically generated visualization is interactive. Users can easily export the generated visualization with the embedded menu.}
\end{figure*}

Data exploration and analysis are critical tasks for organizations of all sizes and industries. The ability to effectively extract insights from a large and complex database can provide a significant competitive advantage and inform strategic decision-making~\citep{provost2013data}. However, for many individuals and organizations, accessing and understanding data can be a daunting task. Traditional data analysis tools often require a significant amount of technical expertise (\eg SQL for querying; Python or R for analysis and visualization) and can be difficult for non-technical professionals to use. 

In this paper, we introduce Mirror, an open-source platform for data exploration and analysis powered by large language and code models, GPT-3~\citep{gpt3} and Codex~\citep{codex}, motivated by prior research on semantic parsing~\citep{zelle1996learning,zettlemoyer2005learning}, data summarization~\citep{ahmed2019data}, and automatic visualization~\citep{zhu2020survey}. The user interface of Mirror is shown in Figure~\ref{fig:mirrorui}. Named after the magic mirror in the fairy tale \textit{Snow White}~\citep{grimm1991snow}, Mirror offers an intuitive natural language interface for querying databases, and automatically generates executable SQL commands to retrieve relevant data and summarize it in natural language.  This %innovative 
approach allows users to easily extract relevant information without having to write complex SQL commands, making data exploration more accessible and efficient. In addition to its natural language querying capabilities, Mirror also allows users to preview and manually edit the generated SQL commands to ensure the accuracy of their queries. This feature provides users with greater control over the data they are extracting, allowing them to fine-tune their queries and ensure that they are getting the most relevant information. To further aid in understanding the data, Mirror also generates visualizations that provide a clear and concise representation of the information. Since pretrained large code and language models like GPT-3 and Codex are good at zero-shot and few-shot inference~\citep{gpt3}, there is no need to collect and annotate supervised data for a new database, making Mirror a ``plug-and-play'' tool for data exploration.

Designed with flexibility and human input in mind, Mirror is suitable for both experienced data analysts and non-technical professionals looking to gain insights from their data and make informed decisions. We further demonstrate two use cases, an automatic question answering application for sports with real-time updates, and OSS Insight Data Explorer, an industrial integration of Mirror for open-source event analysis.

\section{Related Works}
\noindent\textbf{Semantic parsing}~\citep{zelle1996learning,zettlemoyer2005learning} is an important task that converts natural language to a logical form that can be executed by computers. As one major application %one of the most widely acknowledged applications 
of semantic parsing, Text-to-SQL aims to generate an executable program based on the user query. These methods have achieved success when trained and tested on a specific dataset or domain~\citep{t2sql1,t2sql2,t2sql3,t2sql4,spider,dayasql1,dayasql2,t2sql5}. However, how to design a robust and adaptable system that has the ability to plug-and-play for any data source remains unexplored until lately. %an unexplored research question.

Recently, pretrained language models have shown great potential in zero-shot and few-shot inference~\citep{gpt3,t0,flan,chung2022scaling}. Some concurrent works also explored pretrained language models for data querying. \citet{rajkumar2022evaluating} tested zero-shot and few-shot performance of GPT-3~\citep{gpt3} and Codex~\citep{codex} on the Spider~\citep{spider} benchmark.  Without any training, Codex has shown near state-of-the-art performance that is comparable with a fine-tuned T5 model~\citep{t5}. Notably, this work serves as a proof-of-concept and lays a solid foundation for our work.  Binder~\citep{madaan2022language} is a pipeline for few-shot data querying that generates SQL by utilizing the Codex API to fill in designated slots.
% Binder first generates code with slots that call the Codex API to fill.
% After completing the calls and getting the slots filled, the program interpreter (e.g., executor) will fetch the answer for the query. Their experiments show that this pipeline outperforms end-to-end QA or using text-to-SQL with a Codex model on structured data QA datasets. 
Note that different from this work, Mirror focuses on building a human-in-the-loop system for data querying and analysis. Thus, Mirror is orthogonal to Binder and is designed to be generic and compatible with different prompting techniques with minimal modification. 

\begin{figure*}[t]
  \centering
  \includegraphics[width=0.84\linewidth]{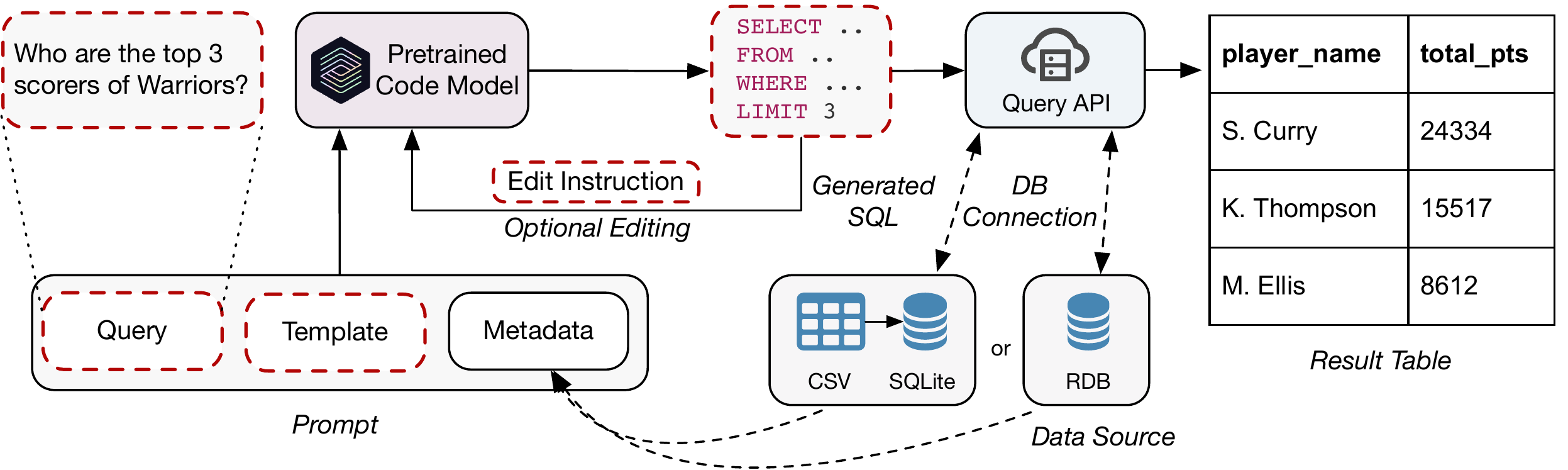}
  \caption{\label{fig:query}The components and workflow of Mirror in the data query stage. The \textcolor{red}{red} dashed lines indicate human-in-the-loop components that can allow user interaction or editing. Query auto-completion with metadata is omitted.}
\end{figure*}

\noindent\textbf{Business intelligence (BI) Tools} are a type of data analysis applications that have user-friendly interfaces which allow people to connect to and query data sources, build dashboards to visualize data and share the results. An example is Redash\footnote{\url{https://redash.io/}}, which uses SQL to retrieve data and supports various visualization formats. Metabase\footnote{\url{https://www.metabase.com/}} provides a graphical user interface for users who cannot write SQL queries as an alternative. Different from these tools, Mirror uses a natural language interface, which further lowers the barrier for data exploration.
Mirror also achieves a higher level of automation that can complete the whole process of data querying, summarization and visualization. %yet allow human interaction between steps. 

\section{System Implementation}
We introduce Mirror in two stages, namely \textit{data query} and \textit{summarization and visualization}. The second stage, \textit{summarization and visualization}, will be automatically triggered every time the data source is successfully queried and returns a result table. Note that all API calls, including OpenAI Codex~\citep{codex} and GPT-3~\citep{gpt3} can be easily substituted with another API service (e.g., Hugging Face API~\citep{hf}) or self-hosted models. For simplicity, we refer to the pretrained code generation model as Codex and text generation model as GPT-3.

\subsection{Data Query}
The components and workflow of the data query stage are shown in Figure~\ref{fig:query}. 

\paratitle{Data Source} When initializing Mirror, a database connection is created by the server-side backend, powered by Node.js. It supports most types of relational databases. Notably, we also enable querying a comma-separated value (CSV) table by converting it to a temporary SQLite instance on the server.

\paratitle{Prompt Construction} There are three components of a prompt that are fed to the pretrained code model. (1) \textit{Template} is a database-specific format string designed by the user. It has slots for \textit{metadata} and \textit{query}. It can also include additional instructions, for example, which SQL dialect to use, alias for table names, value mapping (e.g., airport code to name lookup), or anything that is helpful to achieve better performance on a dataset.
% We also provide a default template (as shown in Appendix~\tba) that is good enough for most use cases. 
(2) \textit{Metadata} are the database schema that can be automatically read from the data source. It is inserted to the corresponding slot in the template. (3) \textit{Query} is the user input in the form of natural language. Mirror also provides  data-aware query auto-completion that completes the user query while typing with the metadata.

\paratitle{SQL Generation} The Codex API is requested after we construct the prompt and it returns a SQL snippet that should be executable. Upon getting the SQL command, a backend query API will be requested to fetch data from the data source. Note that for better security, the database connection should be read-only to prevent SQL injection or accidental alteration (although unlikely). Mirror will retry calling the Codex API if the SQL execution is not successful.\footnote{The Codex API uses random sampling~\citep{topp} to generate code, thus can return different results when retried.} If the API successfully fetches the data from the data source, the subsequent summarization and visualization will be triggered (to be detailed shortly).

\paratitle{Raw and Instructed SQL Editing} We provide a SQL editor with auto-completion that allows users to directly edit the SQL to add human expertise. For non-speakers of SQL, if they are not satisfied with the returned results, they can use natural language instruction to edit the SQL commands. For example, in the example shown in Figure~\ref{fig:query}, if a user wants to exclude retired players from the resulted table and subsequent analysis, they can simply add an instruction of ``Exclude players who have retired``.  This feature is powered by the OpenAI Codex edit API.\footnote{\url{https://beta.openai.com/docs/api-reference/edits}}

\begin{figure*}[t]
  \centering
  \includegraphics[width=0.86\linewidth]{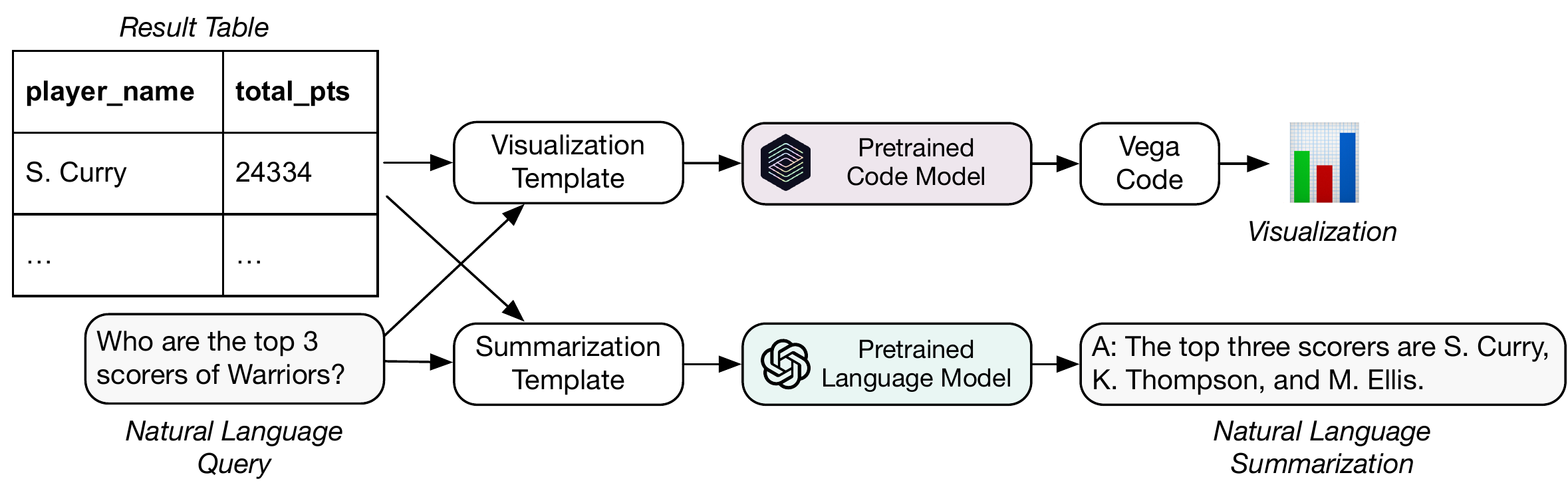}
  \caption{\label{fig:analysis}The components and workflow of Mirror in summarization and visualization stages.}
\end{figure*}

\subsection{Summarization and Visualization}
As shown in Figure~\ref{fig:analysis}, we use a summarization template to organize the query and results into a prompt and use a pretrained language model, GPT-3~\citep{gpt3}, to generate an answer in text form to the query.
% The default summarization template is shown in Appendix~\tba.
Meanwhile, we format the prompt for visualization with a template and the result table as the input. We ask the Codex model to generate a Vega description. Vega\footnote{\url{https://vega.github.io/}} is a declarative language for creating interactive visualization designs. It supports various types of charts, including bar charts, line and area charts, pie charts, scatter plots, distributions and even geographic maps. This enables Mirror to generate a wide range of visualizations for any dataset and purpose. The generated Vega description is in JSON format and can be rendered by the Vega runtime. Similar to SQL generation, if the Vega code fails to compile, the generation will be retried.

\section{Case Study}
\subsection{Automatic Question Answering for Sports}
An automatic question answering system for sports data can be useful for sports fans, media, and managers. Sports data are updated in real time. Thus, large language model-based question answering system (\eg ChatGPT~\citep{chatgpt}) cannot handle such frequent updates. Mirror can be connected to a real-time sports database to provide quick and accurate information to the fans, allowing them to stay informed and engaged with the latest developments in the sports world. It can also be a valuable tool for researching and reporting on sports stories, providing them with up-to-date data to inform the readers and viewers. For managers, this system can provide valuable insights into their team and players through tables and charts, helping them make more informed decisions about strategies and tactics.

\subsection{OSS Insight Data Explorer}
OSS Insight\footnote{\url{https://ossinsight.io/}} is a tool that provides comprehensive, valuable, and trending insights into the open source world by analyzing billions of GitHub events (\eg commit, repository creation, pull requests). It provides in-depth analysis of individual GitHub repositories and developers, as well as the ability to compare two repositories using the same metrics. OSS Insight is a great platform to discover new trends in the open-source community, and empowering new investment and collaboration opportunities.

OSS Insight Data Explorer\footnote{\url{https://ossinsight.io/explore}\label{fn:explorer}} is an industrial application of Mirror. The integration that enables users to query data with Mirror's natural language interface. Mirror enables the users, including non-SQL speakers, e.g., investors, to obtain data, insights, and charts at ease. The live demo is publicly available.\textsuperscript{\ref{fn:explorer}}

\section{Limitations}
\paragraph{Out-of-domain Input} A limitation of Mirror is the handling of out-of-domain input. For example, when a user tries to query the salary of a singer from a sports database, Mirror may generate random or unexecutable SQL, instead of telling the user there is no such information in the database. Thus, it may be useful to allow the generation model to access the SQL execution engine and enable multi-round queries. This may be feasible with a dialogue-based language model (e.g., ChatGPT~\citep{chatgpt}).
\paragraph{Security} Mirror is designed as a business intelligence (BI) tool, thus is mainly for internal use or public data. %It is known 
Previous study~\citep{peng2022security} shows that SQL injection through text-to-SQL models is possible. Thus, we recommend granting Mirror a read-only access in case the generated SQL accidentally modifies the database. Also, access to Mirror should only be granted to a person who is supposed to have full read permission to the database.

\section{Conclusion and Future Work}
In this paper, we propose Mirror, a language model-powered open-source platform for data analysis. With its human-in-the-loop design and zero-shot ability, Mirror is a useful tool for data exploration and analysis that is suitable for many applications. For future work, we will explore dialogue-based multi-round query by combining ChatGPT~\citep{chatgpt} and the SQL engine, to further improve Mirror by improving its self-correction ability and enable more user interaction through a dialogue-like interface. 

\begin{acks}
Mirror was a participating and award-winning project at PingCAP TiDB Hackathon 2022. We would like to thank PingCAP, Inc. for the generous awards and support.
The intellectual property of OSS Insight (including but not limited to its Mirror integration) is owned by PingCAP, Inc. and is not part of the open-source project Mirror described in this paper.
\end{acks}

\bibliographystyle{ACM-Reference-Format}
\bibliography{shrinked}

\appendix









\end{document}